\documentclass[aps, prl, twocolumn, superscriptaddress, nofootinbib, tightenlines]{revtex4-1}

\usepackage{bm,amssymb,slashed,graphicx,multirow,soul,mathtools,xspace,array,booktabs}

\usepackage{hyperref}
\DeclareMathAlphabet{\mathpzc}{OT1}{pzc}{m}{it}

\newcommand{\cbar}{\bar{c}}

\newcommand{\RL}{R(\Lambda_c)}
\newcommand{\Bbar}{\,\overline{\!B}}
\def\B0bar{\Bbar{}^0}

\def\LbarL{\bar\Lambda_\Lambda}
\def\epsc{\varepsilon_c}
\def\epsb{\varepsilon_b}

\def\spnt{1}
\if\spnt1

\else

\fi	

\newcommand{\aS}{\alpha_s}
\newcommand{\haS}{{\hat{\alpha}_s}}
\newcommand{\nn}{\nonumber}
\newcommand{\GeV}{\text{GeV}}

\def\lqcd{\Lambda_\text{QCD}}

\newcommand{\beq}{\begin{equation}}
\newcommand{\eeq}{\end{equation}}
\newcommand{\beqa}{\begin{eqnarray}}
\newcommand{\eeqa}{\end{eqnarray}}

\newcommand{\rC}{r}

\def\d{{\rm d}}
\newcommand{\rl}{\rho_\ell}

\newcommand{\mSqq}{\hat q^2}

\makeatletter
\g@addto@macro\bfseries{\boldmath}
\makeatother

\allowdisplaybreaks

\begin{document}

\title{New predictions for \texorpdfstring{$\Lambda_b\to\Lambda_c$}{LbLc}  semileptonic decays and
tests of heavy quark symmetry}

\author{Florian U.\ Bernlochner}
\affiliation{Karlsruher Institute of Technology, 76131 Karlsruhe, Germany}

\author{Zoltan Ligeti}
\affiliation{Ernest Orlando Lawrence Berkeley National Laboratory, 
University of California, Berkeley, CA 94720, USA}

\author{Dean J.\ Robinson}
\affiliation{Ernest Orlando Lawrence Berkeley National Laboratory, 
University of California, Berkeley, CA 94720, USA}
\affiliation{Santa Cruz Institute for Particle Physics and
Department of Physics, University of California Santa Cruz,
Santa Cruz, CA 95064, USA}

\author{William L.\ Sutcliffe} 
\affiliation{Karlsruher Institute of Technology, 76131 Karlsruhe, Germany}

\begin{abstract}

The heavy quark effective theory makes model independent predictions for semileptonic $\Lambda_b \to \Lambda_c$ decays in terms of a small set of parameters.  No subleading Isgur-Wise function occurs at order $\lqcd/m_{c,b}$, and only two sub-subleading functions enter at order $\lqcd^2/m_c^2$.  These features allow us to fit the form factors and decay rates calculated up to order $\lqcd^2/m_c^2$ to LHCb data and lattice QCD calculations. We derive a significantly more precise standard model prediction for the ratio ${\cal B}(\Lambda_b\to \Lambda_c \tau\bar\nu) / {\cal B}(\Lambda_b\to \Lambda_c \mu\bar\nu)$ than prior results, and find the expansion in $\lqcd/m_c$ well-behaved, addressing a long-standing question.
Our results allow more precise and reliable calculations of $\Lambda_b\to
\Lambda_c\ell\bar\nu$ rates, and are systematically improvable with better data
on the $\mu$ (or $e$) modes.

\end{abstract}

\maketitle

\section{Introduction}

Semileptonic decays mediated by $b\to c\ell\bar\nu$ transitions give tantalizing hints of deviations from the standard model (SM), in the ratios
\beq\label{RXdef}
R(D^{(*)}) = {\Gamma(B\to D^{(*)}\tau\bar\nu)} \big/ {\Gamma(B\to D^{(*)} l\bar\nu)}\,,
\eeq
where $l = \mu, e$. Combining the $D$ and $D^*$ results, the tension with the SM is $4 \sigma$~\cite{Amhis:2016xyh}. Precision control of hadronic matrix elements are crucial to predict the ratios of decay rates: A better understanding of the heavy quark expansion to $\mathcal{O}(\lqcd^2/m_c^2)$ is required, as it is largely responsible for the different uncertainty estimates of $R(D^*)$ in the SM~\cite{Bernlochner:2017jka, Bigi:2017jbd, Jaiswal:2017rve}. The same hadronic matrix elements are also crucial to resolve tensions between inclusive and exclusive determinations of $|V_{cb}|$~\cite{Bernlochner:2017jka, Bigi:2017njr, Grinstein:2017nlq, Bernlochner:2017xyx, Lattice:2015rga, BDsLatticeAllw, Bigi:2017jbd, Jaiswal:2017rve}. 
These anomalies triggered exploring a vast array of models, e.g., with TeV-scale leptoquarks or exotic gauge bosons, as well as new high-$p_T$ searches at the LHC for the possible mediators. 

The $\Lambda_b \to \Lambda_c \ell \bar\nu$ baryon decays provide a theoretically
cleaner laboratory than $B\to D^{(*)} \ell\bar\nu$ to examine $\mathcal{O}(\lqcd^2/m_c^2)$ terms, as heavy quark symmetry~\cite{Isgur:1989vq, Isgur:1989ed, Isgur:1991wq} provides stronger constraints. The $\mathcal{O}(\lqcd/m_{c,b})$ contributions yield no new nonperturbative functions beyond the leading order Isgur-Wise function, significantly reducing the number of hadronic parameters order by order in the heavy quark effective theory (HQET)~\cite{Georgi:1990um, Eichten:1989zv} description of these decays.  
This allows us to determine the $\mathcal{O}(\lqcd^2/m_c^2)$ contributions to an exclusive decay for the first time, without any model dependent assumption.

In this letter we examine the HQET predictions at $\mathcal{O}(\lqcd^2/m_c^2)$ and fit them to a recent LHCb measurement of $\Lambda_b\to \Lambda_c \mu\bar\nu$~\cite{Aaij:2017svr} and/or lattice QCD (LQCD)
results~\cite{Detmold:2015aaa}. Doing so, we obtain the most precise SM
prediction so far for
\beq\label{RLdef}
  \RL = {\Gamma(\Lambda_b\to \Lambda_c\tau\bar\nu)} \big/
    {\Gamma(\Lambda_b\to \Lambda_c \mu\bar\nu)}\,, 
\eeq
improvable with future data.  We find that the $\mathcal{O}(\lqcd^2/m_c^2)$ corrections have the expected characteristic size, suggesting that the heavy quark expansion 
in $\lqcd/m_c$ is well behaved in such decays. 

Testing HQET predictions not only provides a path to reducing
theoretical uncertainties in precision determinations of $R(D^{(*)})$ and the extraction of
$|V_{cb}|$, but also improves the sensitivity to possible new physics
contributions.
Measuring semileptonic decays mediated by the same
parton-level transition between different hadrons is important, as it improves
the statistics, entails different systematic uncertainties, and gives
complementary information on possible new physics.  LHCb projections show that
the precision of $\RL$ will be near those of $R(D^{(*)})$ in the
future~\cite{LHCbprojection}, making this channel very important.

\section{HQET expansion of the form factors}
\label{sec:hqet}

The semileptonic $\Lambda_b\to \Lambda_c\ell\bar\nu$ form factors in HQET are
conventionally defined for the SM currents as~\cite{Isgur:1990pm, Falk:1992ws,
Manohar:2000dt}
\begin{align}\label{HQETffdef}
&\langle \Lambda_c(p',s')| \bar c\gamma_\nu b |\Lambda_b(p,s)\rangle \nn \\*
 & \qquad = \bar u_c(v',s') \big[ f_1 \gamma_\mu + f_2 v_\mu + f_3 v'_\mu \big] u_b(v,s)\,, \nn\\*
&\langle \Lambda_c(p',s')| \bar c\gamma_\nu\gamma_5 b |\Lambda_b(p,s)\rangle \nn \\* 
& \qquad = \bar u_c(v',s') \big[ g_1 \gamma_\mu + g_2 v_\mu + g_3 v'_\mu \big] \gamma_5\, u_b(v,s)\,,
\end{align}
where $p = m_{\Lambda_b}v$, $p' = m_{\Lambda_c}v'$, and the $f_i$ and $g_i$ form factors are functions of $w = v \cdot v'$.  The spinors are normalized to $\bar u u = 2m$.

The $\Lambda_{b,c}$ baryons are singlets of heavy quark spin symmetry, with the
``brown muck" of the light degrees of freedom in the spin-$0$ ground state.  Therefore,
\begin{equation}\label{mass}
m_{\Lambda_Q} = m_Q + \LbarL - \lambda_1^\Lambda / 2 m_Q
  + \ldots \,,
\end{equation}
where $Q=b,c$, the ellipsis denotes terms higher order in $\lqcd/m_Q$ and $m_{\Lambda_b} =
5.620\,$GeV, $m_{\Lambda_c} = 2.286$\,GeV~\cite{PDG}.
The parameter $\LbarL$  is the energy of the light degrees of freedom
in the $m_Q \gg \lqcd$ limit, and $\lambda_1^\Lambda$ is related to
the heavy quark kinetic energy in the $\Lambda_{b,c}$ baryons.
Using a short-distance quark mass scheme, ambiguities in the pole mass and $\LbarL$ can be canceled, and the behavior of the perturbation series improved. We use the $1S$ scheme~\cite{Hoang:1998ng, Hoang:1998hm, Hoang:1999ye} and
treat $m_b^{1S} = (4.71\pm0.05)\,\GeV$ and $\delta m_{bc}
= m_b-m_c = (3.40\pm0.02)\,\GeV$ as independent
parameters~\cite{Ligeti:2014kia, Bernlochner:2017jxt}.  (The latter is well constrained by $B\to
X_c\ell\bar\nu$ spectra~\cite{Bauer:2004ve, Bauer:2002sh}.)  
We match HQET onto QCD at $\mu = \sqrt{m_cm_b}$, so that $\aS \simeq 0.26$. 
For example, using Eq.~(\ref{mass}) for both $\Lambda_b$ and $\Lambda_c$ to eliminate $\lambda_1^\Lambda$, at $\mathcal{O}(\aS)$ we obtain $\LbarL = (0.81 \pm 0.05)\,\GeV$.

Making the transition to HQET~\cite{Georgi:1990um, Eichten:1989zv}, at leading
order in the heavy quark expansion
\begin{equation}\label{leading}
\langle \Lambda_c(p',s')| \bar c\, \Gamma b\, |\Lambda_b(p,s)\rangle 
  = \zeta(w)\, \bar u_c(v',s')\, \Gamma\, u_b(v,s) \,,
\end{equation}
where $u(v,s)$ satisfies $\slashed{v}\, u = u$ and $\zeta(w)$ is the leading order Isgur-Wise function~\cite{Isgur:1990pm}, satisfying $\zeta(1)=1$.  In the heavy quark limit, $f_1 = g_1 = \zeta$, while $f_{2,3} = g_{2,3} = 0$.

At order $\lqcd/m_{c,b}$ a remarkable simplification occurs compared to meson decays:
The $\mathcal{O}(\lqcd/m_{c,b})$ corrections from the matching of the $\cbar\, \Gamma b$ heavy quark current onto HQET~\cite{Falk:1990yz,
Falk:1990cz, Neubert:1992qq} can be expressed in terms of $\LbarL$ and the
leading order Isgur-Wise function $\zeta(w)$~\cite{Georgi:1990ei}. 
In addition, for $\Lambda_b \to \Lambda_c$ transitions, there are no ${\cal O}(\lqcd/m_{c,b})$ contributions from the chromomagnetic operator. 
The kinetic energy
operator in the ${\cal O}(\lqcd/m_{c,b})$ HQET Lagrangian gives rise to 
a heavy quark spin symmetry conserving subleading
term, parametrized by $\zeta_{\rm ke}(w)$, which can be absorbed into the leading order Isgur-Wise function by redefining $\zeta$ via
\begin{equation}\label{eqn:KEabs}
\zeta(w) +  (\epsc + \epsb)\, \zeta_{\rm ke}(w) \to \zeta(w)\,,
\end{equation}
where $\varepsilon_{c,b} = \LbarL/(2\, m_{c,b})$.
Thus, no additional unknown functions beyond $\zeta(w)$ are needed to parametrize the ${\cal O}(\lqcd/m_{c,b})$
corrections. Luke's theorem~\cite{Luke:1990eg} implies $\zeta_{\rm ke}(1) = 0$, so
the normalization $\zeta(1) = 1$ is preserved.
Perturbative corrections to the heavy quark
currents can be computed by matching QCD onto HQET~\cite{Falk:1990yz,
Falk:1990cz, Neubert:1992qq}, and introduce no new hadronic parameters.

The $\mathcal{O}(\lqcd^2/m_{c,b}^2)$ corrections are
parametrized by six unknown functions of $w$~\cite{Falk:1992ws}, but only two linear combinations of sub-subleading Isgur-Wise functions, $b_{1,2}$, occur at
$\mathcal{O}(\lqcd^2/m_c^2)$. Spurious terms introduced by the redefinition in Eq.~\eqref{eqn:KEabs} at order $\lqcd^2/m_c^2$ can also be absorbed into $b_{1,2}$.
We define the rescaled form factors,
\begin{equation}\label{eqn:hatHdef}
\hat x_i(w) = x_i(w) \big/ \zeta(w)\,, \qquad x = \big\{f_i\,,\ g_i\,,\ b_i\big\}\,.
\end{equation}

Including $\alpha_s$, $\lqcd/m_{c,b}$, $\alpha_s\,
\lqcd/m_{c,b}$~\cite{Neubert:1993mb}, and $\lqcd^2/m_c^2$ corrections, the SM
form factors are
\begin{widetext}
\begin{align}
\label{ffexpsm}
\hat f_1 &= 1 + \haS C_{V_1} + \epsc + \epsb + \haS \Big[ C_{V_1}
   + 2(w-1)C'_{V_1} \Big] (\epsc + \epsb)
  + \frac{\hat b_1-\hat b_2}{4m_c^2} + \ldots \,, \nn\\*
\hat f_2 &= \haS C_{V_2} - \frac{2\, \epsc}{w+1}
  + \haS\bigg[ C_{V_2} \frac{3w-1}{w+1} \epsb - \big[2C_{V_1} - (w-1) C_{V_2} + 2C_{V_3}\big]
  \frac{\epsc}{w+1} 
  + 2(w-1)C'_{V_2} (\epsc + \epsb)\bigg]
  + \frac{\hat b_2}{4m_c^2} + \ldots , \nn\\*
\hat f_3 &= \haS C_{V_3} - \frac{2\, \epsb}{w+1} 
  + \haS\bigg[  C_{V_3} \frac{3w-1}{w+1} \epsc
  - \big[2C_{V_1} + 2C_{V_2} - (w-1) C_{V_3}\big] \frac{\epsb}{w+1}
  + 2(w-1) C'_{V_3} (\epsc + \epsb)\bigg] + \ldots \,, \nn\\
\hat g_1 &= 1 + \haS C_{A_1} + (\epsc + \epsb)\, \frac{w-1}{w+1} 
  + \haS\bigg[ C_{A_1}\, \frac{w-1}{w+1}
  + 2(w-1)C'_{A_1} \bigg] (\epsc+\epsb)
  + \frac{\hat b_1}{4m_c^2} + \ldots \,, \nn\\*
\hat g_2 &= \haS C_{A_2} - \frac{2\, \epsc}{w+1}
  + \haS \bigg[C_{A_2} \frac{3w+1}{w+1} \epsb - \big[2C_{A_1} - (w+1) C_{A_2} + 2C_{A_3}\big]
  \frac{\epsc}{w+1}
  + 2(w-1)C'_{A_2} (\epsc + \epsb)\bigg]
  + \frac{\hat b_2}{4m_c^2} + \ldots , \nn\\*
\hat g_3 &= \haS C_{A_3} + \frac{2\, \epsb}{w+1}
  + \haS\bigg[C_{A_3} \frac{3w+1}{w+1} \epsc
  +  \big[2C_{A_1} - 2C_{A_2} + (w+1) C_{A_3}\big] \frac{\epsb}{w+1}
  + 2(w-1)C'_{A_3} (\epsc + \epsb)\bigg] + \ldots \,,
\end{align}
\end{widetext}
where the $C_{\Gamma_i}$ are functions of $w$~\cite{Bernlochner:2017jka,
Neubert:1992qq}, $z=m_c/m_b$, and $\haS = \aS/\pi$.  (We use the notation of
Ref.~\cite{Manohar:2000dt}; explicit expressions for $C_{\Gamma_i}$ are in Ref.~\cite{Bernlochner:2017jka}.)  In
Eq.~(\ref{ffexpsm}), a prime denotes $\partial/\partial w$
and the ellipses denote ${\cal O}(\epsc\epsb,\,
\epsb^2,\, \epsc^3)$ and higher order terms.
Equation~\eqref{ffexpsm} agrees with Eq.~(4.75) in Ref.~\cite{Neubert:1993mb} (where a different form of Eq.~\eqref{eqn:KEabs} is used).

The $\hat b_{1,2}(w)$ functions
are not constrained by heavy quark symmetry.  The model dependent estimate
$\hat b_1(1) \approx -3\LbarL^2$, obtained in Eq.~(5.5) of
Ref.~\cite{Falk:1992ws}, would imply that $\hat b_1/(4m_c^2)$ terms can
give ${\cal O}(20\%)$ corrections.  Even corrections of such size would not
necessarily imply a breakdown of the heavy quark expansion:  A matrix element $\sim 3\LbarL^2$ is consistent with HQET power counting, as dependence of the form factors on
the energy of the brown muck in the hadron, $\LbarL$, arises from using the equations of
motion.  Since $\LbarL$ is greater than $\bar\Lambda$ in the $B \to D^{(*)}$ case~\cite{Bernlochner:2017jka}, it would not be surprising if 
the HQET expansions for $\Lambda_b \to \Lambda_c$ form factors converge slower than for $B \to D^{(*)}$.  
At the same time, the structure of the expansion is simpler for $\Lambda_b\to \Lambda_c$
form factors (cf.\ similar HQET-based discussions of $B\to
D^{(*)}\ell\bar\nu$~\cite{Bernlochner:2017jka}, $B\to
D^{**}\ell\bar\nu$~\cite{Leibovich:1997tu, Leibovich:1997em,
Bernlochner:2016bci, Bernlochner:2017jxt}, and $\Lambda_b \to
\Lambda_c^*\ell\bar\nu$~\cite{Leibovich:1997az, Boer:2018vpx}).

\section{Fits to LHC\lowercase{b} and lattice QCD data}

To determine the nonperturbative quantities that occur in the HQET expansion of
the form factors in Eq.~(\ref{ffexpsm}), assess the behavior of the expansion in
$\lqcd/m_c$, and derive precise SM predictions for $\RL$ in Eq.~(\ref{RLdef}),
we fit the LHCb measurement of $\d\Gamma(\Lambda_b\to \Lambda_c
\mu\bar\nu)/\d q^2$~\cite{Aaij:2017svr} or/and a LQCD determination of
the six form factors~\cite{Detmold:2015aaa}. 

\begin{figure}[t]
\includegraphics[width=0.95\columnwidth, clip, bb=10 0 420 310]{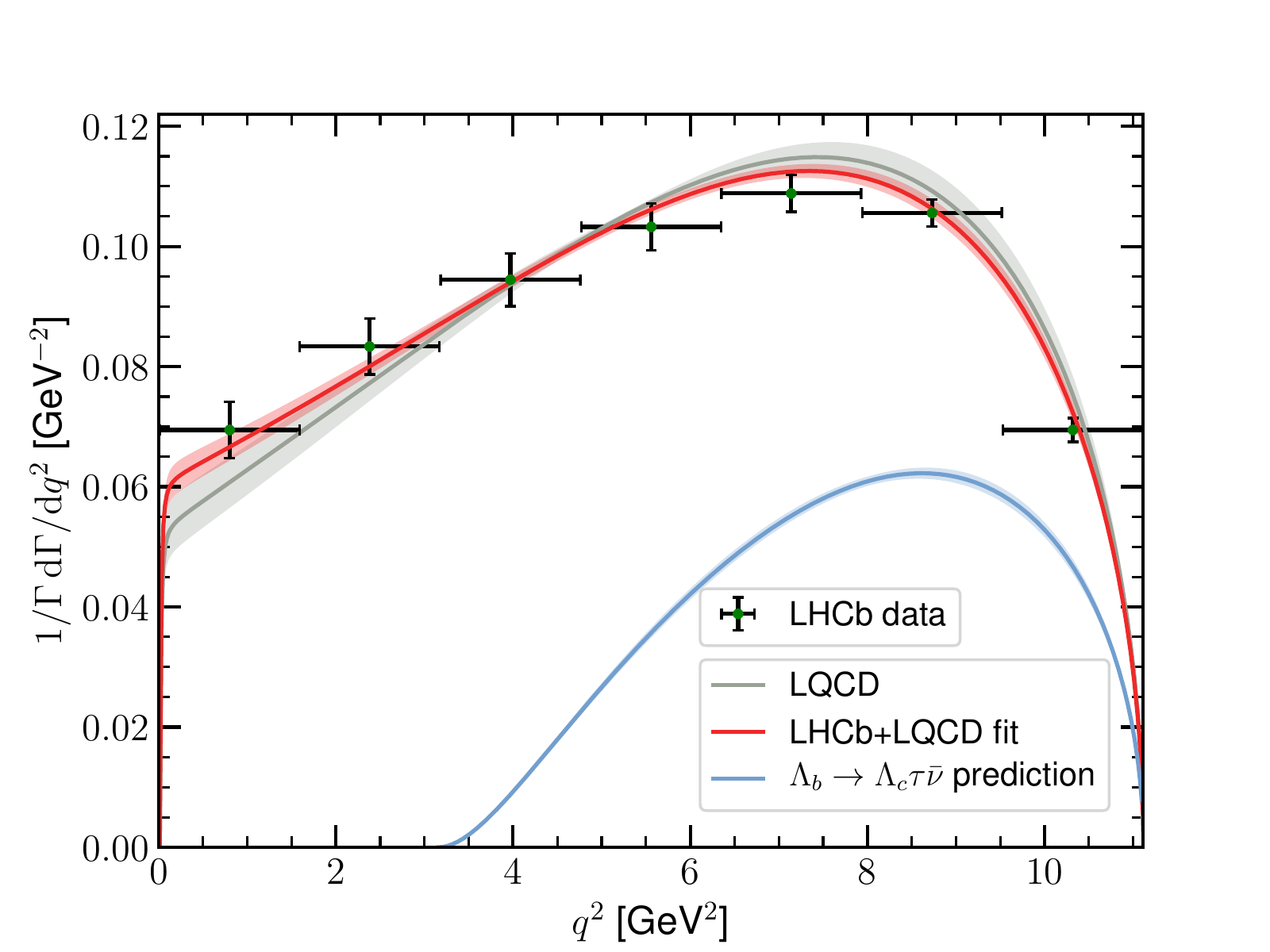}
\caption{The red band shows our fit of the HQET predictions to
$\d\Gamma(\Lambda_b\to \Lambda_c \mu\bar\nu)/\d q^2$ measured by LHCb~\cite{Aaij:2017svr} and the LQCD form factors~\cite{Detmold:2015aaa}.  The heather gray band shows the LQCD prediction.  The blue curve shows our prediction for $\d\Gamma(\Lambda_b\to
\Lambda_c \tau\bar\nu)/\d q^2$.}
\label{fig:q2spec}
\end{figure}

The LHCb experiment measured the $q^2$ spectrum in 7 bins, normalized to unity~\cite{Aaij:2017svr}.  This reduces its effective degrees of freedom from 7 to 6 (as any one bin is determined by the sum of the others).  The measurement is shown as the data points in Fig.~\ref{fig:q2spec}. 

\begin{figure*}[t]
\includegraphics[width=0.33\textwidth, clip, bb=0 0 420 315]{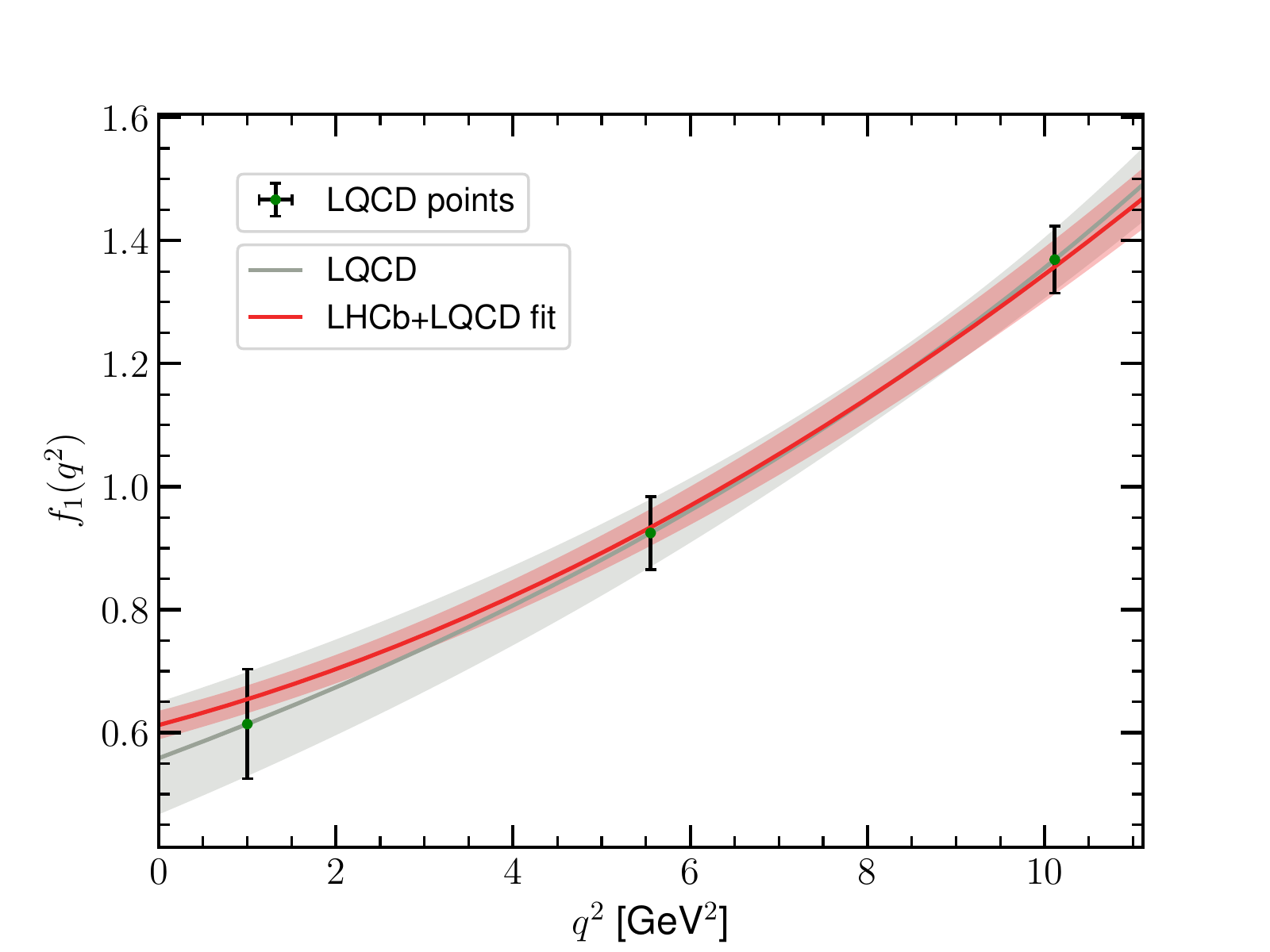}\hfill
\includegraphics[width=0.33\textwidth, clip, bb=0 0 420 315]{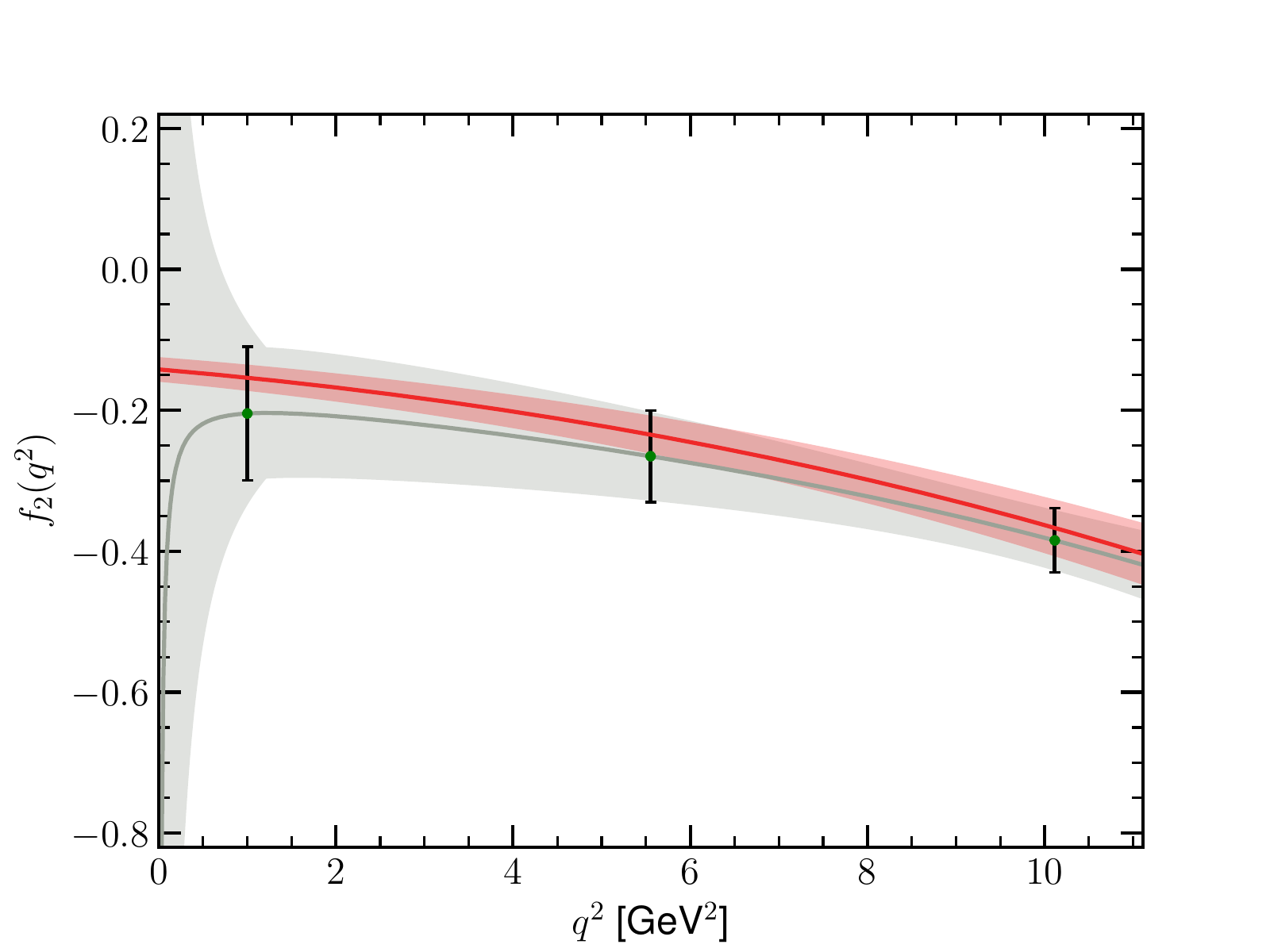}\hfill
\includegraphics[width=0.33\textwidth, clip, bb=0 0 420 315]{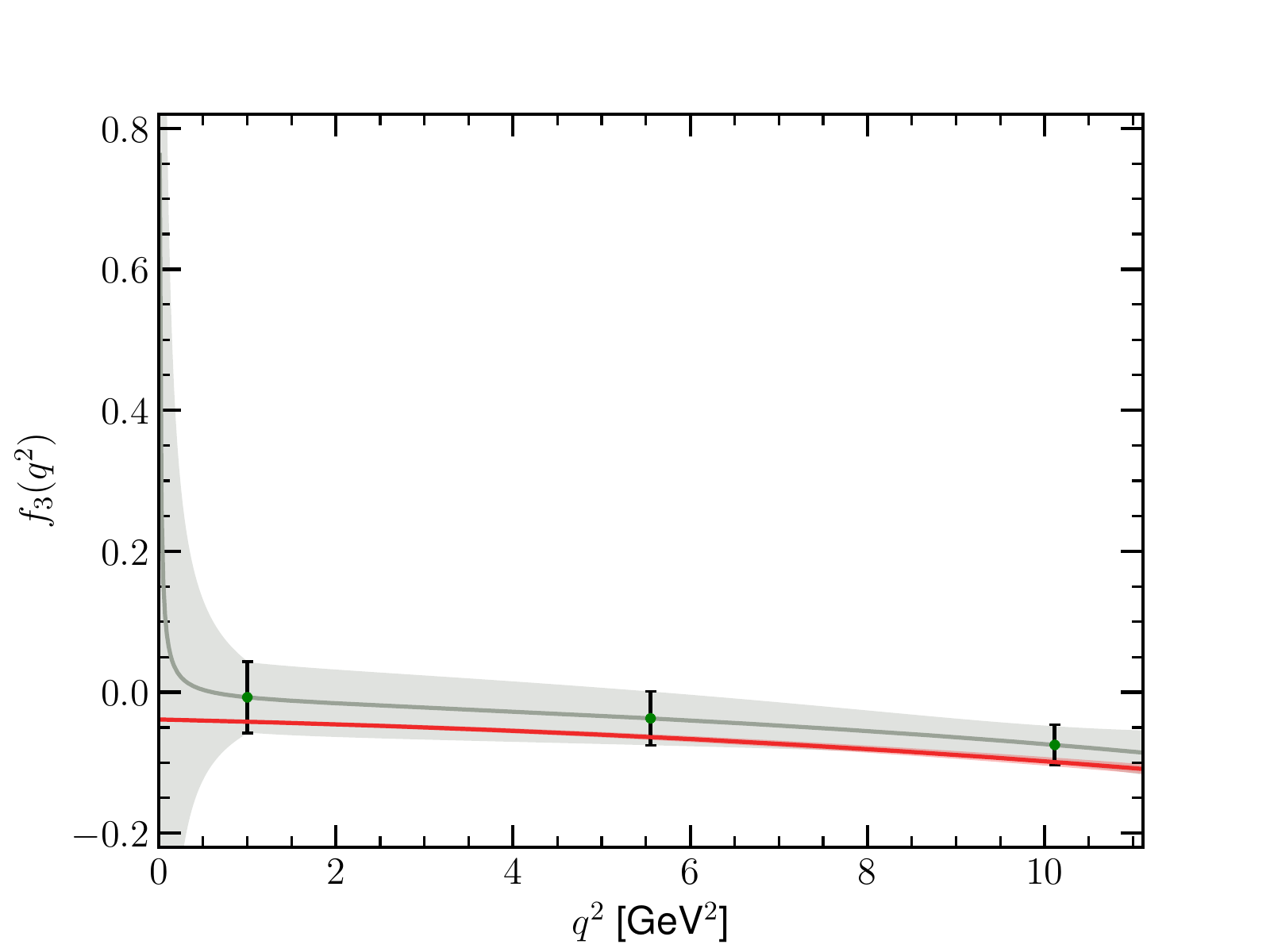}
\\[6pt]
\includegraphics[width=0.33\textwidth, clip, bb=0 0 420 315]{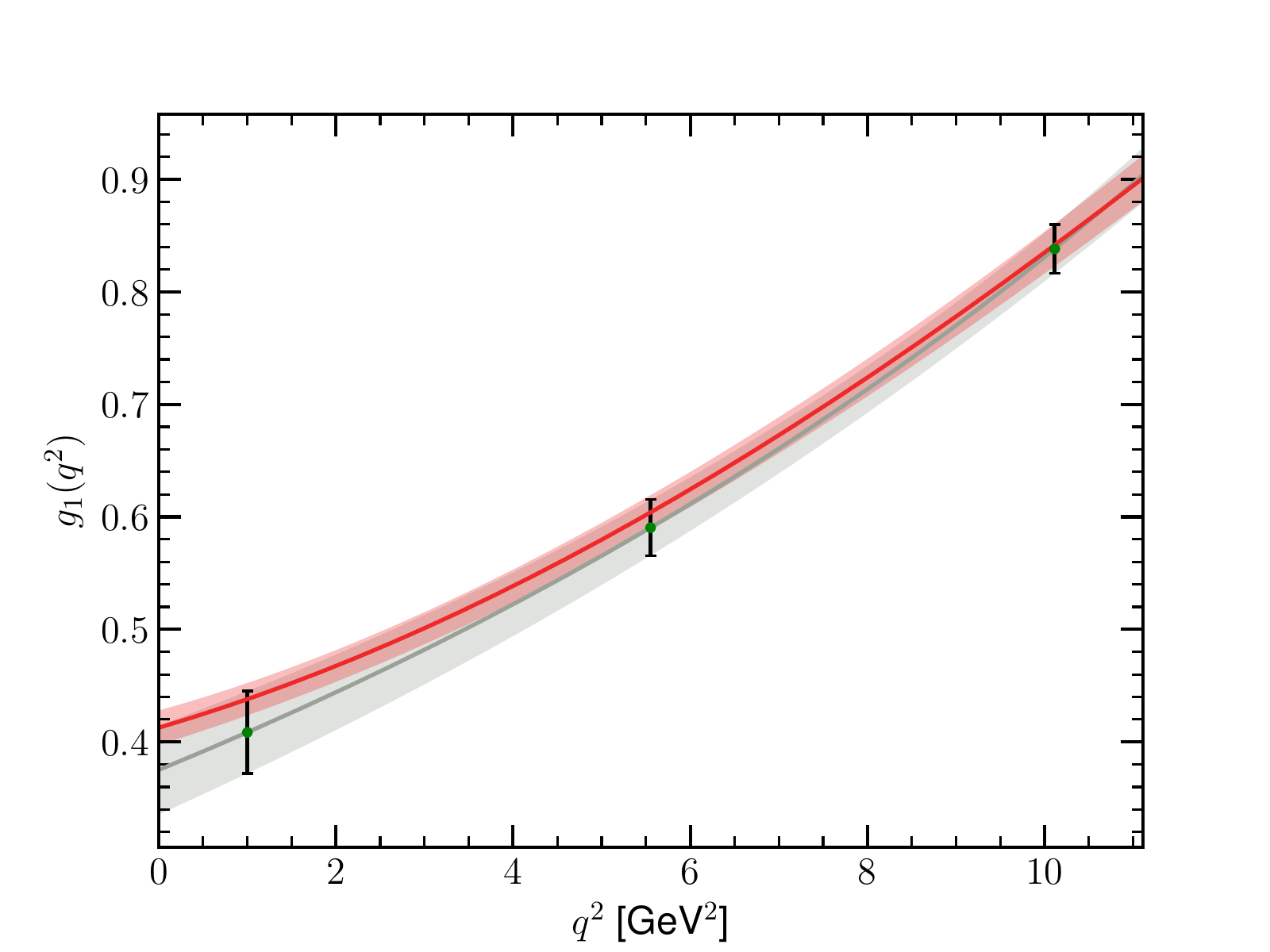}\hfill
\includegraphics[width=0.33\textwidth, clip, bb=0 0 420 315]{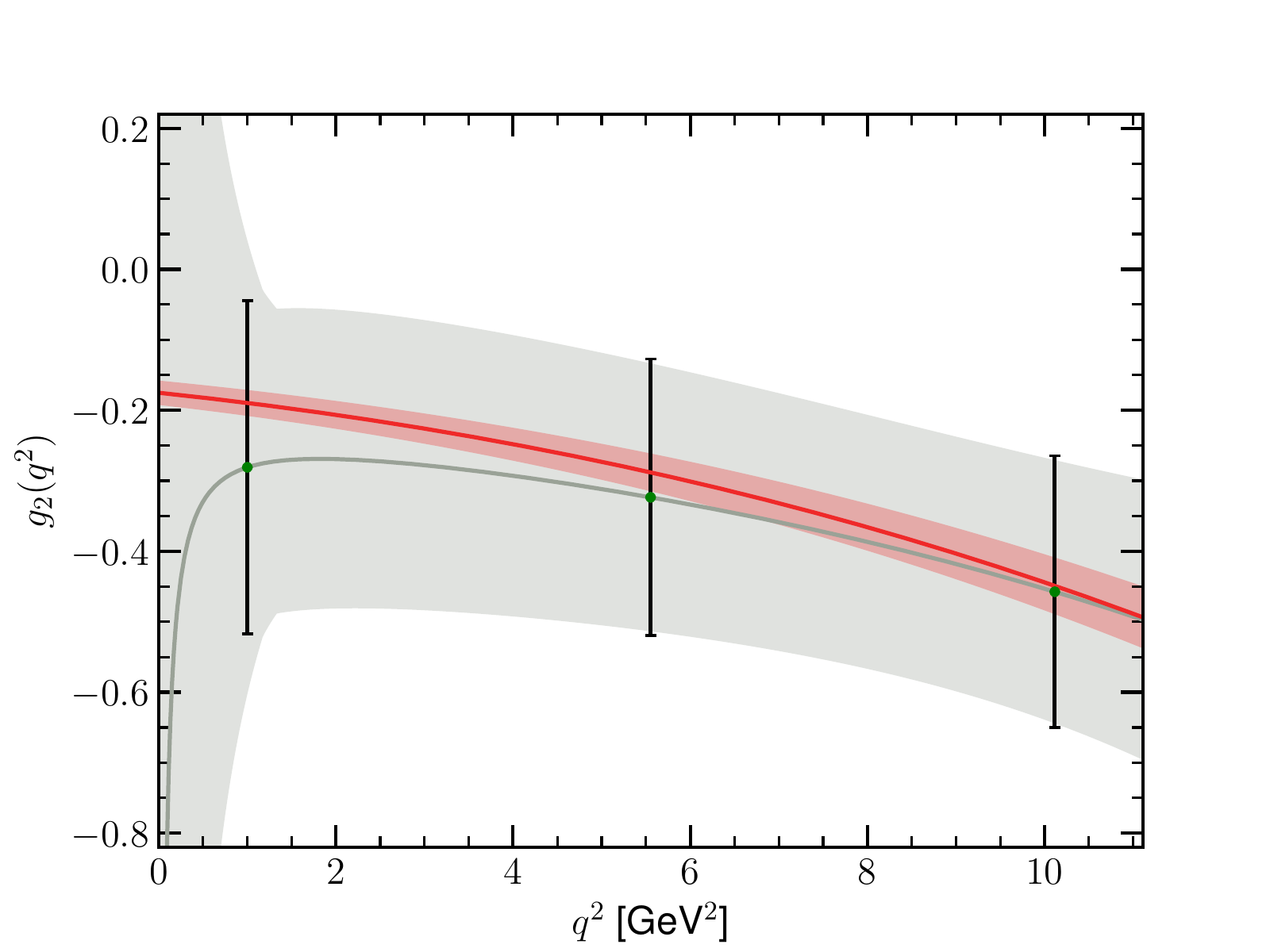}\hfill
\includegraphics[width=0.33\textwidth, clip, bb=0 0 420 315]{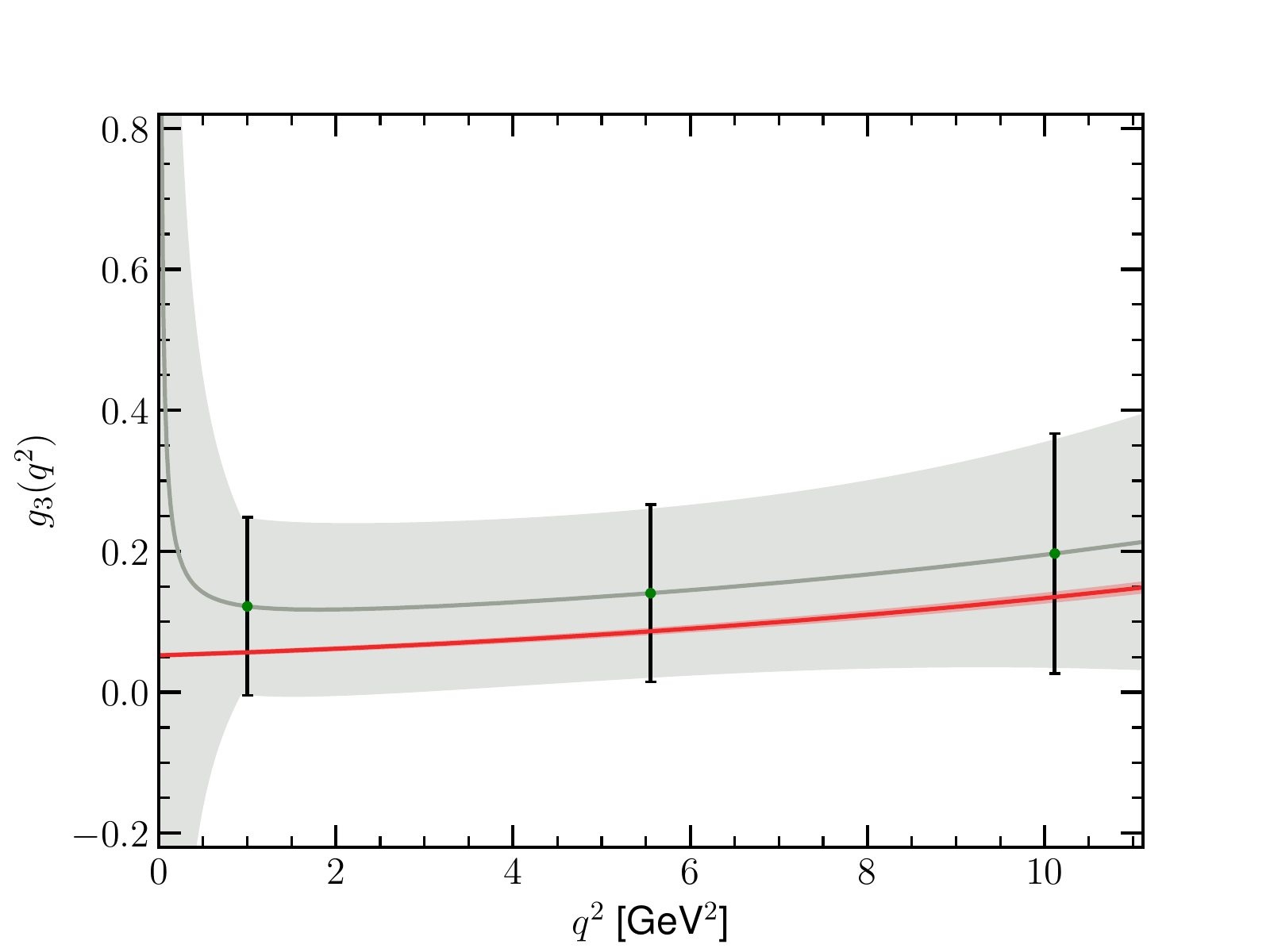}
\caption{Fit of the HQET predictions in Eq.~(\ref{ffexpsm}) to the LQCD
results~\cite{Detmold:2015aaa} and the LHCb spectrum~\cite{Aaij:2017svr} for the 6 form factors (red bands). The heather gray bands and data points show the LQCD prediction; see text for details.}
\label{fig:lqcd_fits}
\end{figure*}

The lattice QCD results~\cite{Detmold:2015aaa} for the 6 form factors are published as fits to the BCL parametrization~\cite{Bourrely:2008za}, using either 11 or 17 parameters. We derive predictions for $f_{1,2,3}$ and $g_{1,2,3}$ using the 17 parameter result at three $q^2$ values, near the two ends and the middle of the spectrum, $q^2 = \big\{ 1 \, \text{GeV}^2, \ q^2_{\rm max}/2, \ q^2_{\rm max} - 1 \, \text{GeV}^2\big\}$, preserving their full correlation, in order to construct an appropriate covariance matrix. The difference in the form factor values obtained using the 17 or the 11 parameter results is added as an uncorrelated uncertainty. This differs slightly from the prescription in Ref.~\cite{Detmold:2015aaa}, based on the maximal differences, which cannot preserve the correlation structure between the form factor values. The 18 form factor values used in our fits are shown as data points in Fig.~\ref{fig:lqcd_fits}. The LQCD predictions, following the prescription of Ref.~\cite{Detmold:2015aaa}, are shown as heather gray bands, and the uncertainties are in good agreement.  The heather gray band in Fig.~\ref{fig:q2spec} shows the LQCD prediction for the normalized spectrum, using the BCL parametrization.

The SM prediction for the decay rate for arbitrary charged lepton mass is
\begin{align}\label{dGdwSM}
\frac{\d\Gamma}{\d w} 
	& = \frac{G_F^2\, m_{\Lambda_b}^5 |V_{cb}|^2}{24\,\pi^3}\frac{(\mSqq - \rl)^2}{\hat q^4}\, \rC^3\, \sqrt{w^2-1}\,
	\bigg\{\! \bigg(1 + \frac{\rl}{2\mSqq}\bigg) \nn \\*
	& \times \Big[ (w-1)\big(2\mSqq f_1^2 + \mathcal{F}_+^2\big) + (w+1)\big(2\mSqq g_1^2 + \mathcal{G}_+^2\big) \Big] \nn\\*
	& + \frac{3\rl}{2\mSqq} \Big[(w+1) \mathcal{F}_0^2 + (w-1) \mathcal{G}_0^2\Big]\bigg\}\,,
\end{align}
where $\rl = m_\ell^2/m_{\Lambda_b}^2$, $\rC = m_{\Lambda_c} /
m_{\Lambda_b}$, $\mSqq \equiv q^2/m_{\Lambda_b}^2 = 1 - 2\rC w + \rC^2$, and
\begin{align}
\label{eqn:FFrel}
\mathcal{F}_+ & = (1+\rC)f_1 + (w+1)(\rC\, f_2+f_3) = (1+\rC) f_+\,, \, \\*
\mathcal{G}_+ & = (1-\rC)g_1 - (w-1)(\rC\, g_2+g_3) = (1-\rC) g_+\,, \nn\\
\mathcal{F}_0 & = (1-\rC)f_1 - (\rC w - 1) f_2 + (w-\rC)f_3 = (1-\rC) f_0\,, \nn\\*
\mathcal{G}_0 & = (1 + \rC)g_1 + (\rC w - 1)g_2 - (w-\rC)g_3 = (1+\rC) g_0\,.\nn
\end{align}
Combined with $f_1 = f_{\perp}$ and $g_1 = g_{\perp}$, Eqs.~\eqref{eqn:FFrel} relate $f_i$ and $g_i$ to the other common form factor basis, $f_{\perp, +, 0}$ and $g_{\perp, +, 0}$, used in Ref.~\cite{Detmold:2015aaa}. 
Our result in Eq.~\eqref{dGdwSM} agrees with those in Refs.~\cite{Boyd:1997kz, Detmold:2015aaa}.

In our fits to the LHCb data, we integrate the rate predictions that follow from
Eqs.~(\ref{ffexpsm}) and (\ref{dGdwSM}) over each bin, and minimize a $\chi^2$ function.  The LQCD predictions are fitted by minimizing a $\chi^2$ function that includes the 18 values and their correlations, as described above.

\begin{table}[b]
\centerline{\scalebox{0.95}{
\renewcommand{\arraystretch}{1.3}
\begin{tabular}{c|ccc}
\hline\hline
& LHCb   & LQCD & LHCb + LQCD \\
\hline 
$\zeta'$ & $ -2.17 \pm 0.26$  					& $ -2.05 \pm 0.13$	  	&  $ -2.04 \pm 0.08$ 	\\
$\zeta''$ &$ 4.10 \pm 1.05$ 				        & $ 2.93 \pm 0.43$		&  $ 3.16 \pm 0.38$ \\
$\hat b_1$/GeV${}^2$ &$ 0.24 \pm 1.92\, {}^*$ 				& $ -0.44 \pm 0.16$	 	& $ -0.46 \pm 0.15$	\\
$\hat b_2$/GeV${}^2$ & $ 0.45 \pm 1.88\, {}^*$				& $ -0.41 \pm 0.40$	 	& $ -0.39 \pm 0.39$  \\ \hline
$m_b^{1S}$/GeV &$ 4.71 \pm 0.05$		& $ 4.72 \pm 0.05$		& $ 4.72 \pm 0.05$   \\
$\delta m_{bc}$/GeV &$ 3.40 \pm 0.02$		& $ 3.40 \pm 0.02$		& $ 3.40 \pm 0.02$   \\
\hline
$\chi^2/\text{ndf}$ & 0.77/4 & $ 2.42 / 14$ & $ 7.20 / 20$  \\ 
\hline
$\RL$ &  $0.3209 \pm 0.0041$& $0.3313 \pm 0.0101$ & {\bf $0.3237 \pm 0.0036$} \\
\hline\hline
\end{tabular}
}}
\caption{HQET parameters extracted from the 3 fits discussed in the text.  Predictions for $\RL$ for each
fit are shown in the last row. The $\hat b_{1,2}$ values marked with an asterisk were constrained in the fit; see text for details. }
\label{tab:fitsummary}
\end{table}

We explore three scenarios: (i) fitting only the LHCb spectrum; (ii) fitting
only the LQCD data; and (iii) a combined fit the the LHCb data and the
LQCD information.  The resulting HQET parameters are summarized in
Table~\ref{tab:fitsummary}. For the fit to only the LHCb spectrum, the unknown
absolute normalization of the measurement removes the sensitivity to $\hat
b_{1,2}$.  Therefore, we constrain them to 0 by a Gaussian with a $2\,\GeV^2
(\approx 3\LbarL^2)$ uncertainty, motivated by a model dependent estimate for
$\hat b_1(1)$~\cite{Falk:1992ws}. This allows our 3 fits to have the same
parameters, and be compared to one another. In all fits, $m_b^{1S}$ and $\delta
m_{bc}$ are constrained using Gaussian uncertainties.  The leading order Isgur-Wise
function is fitted as
$\zeta = 1 + (w-1) \zeta' + \frac12 (w-1)^2 \zeta''$. Alternative
expansions using the conformal parameters $\mathpzc{z}$ or $\mathpzc{z}^*$ instead of $w$ yield
nearly identical fits. Fits with $\zeta$ linear in either $w$, $\mathpzc{z}$, or $\mathpzc{z}^*$ are poor.
Adding more $q^2$ values from the BCL fit of the LQCD result to our sampling indicates no preference for the inclusion of higher order terms in $w-1$, nor does it noticeably affect the fit results.
We fit $\hat b_{1,2}$ as constants, which is appropriate at the current level of sensitivity.
We do not include explicitly an uncertainty for neglected higher order terms in Eq.~\eqref{ffexpsm}; two form factors, $f_3$ and $g_3$, receive no $\lqcd^2/m_c^2$ corrections, so their agreement with LQCD in the right-most plots in Fig.~\ref{fig:lqcd_fits} indicates that these terms are probably small.

All fits have acceptable $\chi^2$ values, and they all yield compatible values
for the slope and curvature of $\zeta(w)$ at zero recoil. The fit of the
HQET predictions to the lattice QCD form factors determines fairly precisely the
$\hat b_{1,2}$ parameters, that enter at order
$\lqcd^2/m_c^2$.  The significance of $\hat b_1 \neq 0$ is over
$3\sigma$.  However $\hat b_1(1)$ is much smaller than the model dependent
estimate $\hat b_1(1) \simeq -3\LbarL^2$~\cite{Falk:1992ws}.

The red bands in Figures~\ref{fig:q2spec} and ~\ref{fig:lqcd_fits} show the
combined fit using both LHCb and LQCD information. The agreement therein shows that the HQET predictions in Eq.~(\ref{ffexpsm} describe the
form factors and the experimental spectrum at the current level of uncertainties.
This also holds for the fit using the LHCb spectrum (with constraints on $\hat b_{1,2}$). Table~\ref{tab:fitcov} shows the correlation matrix of the LHCb + LQCD
fit. 
Table~\ref{tab:fitsummary} also shows the resulting SM predictions for $\RL$ from the 3 fits, and Fig.~\ref{fig:q2spec} shows the predicted $\d\Gamma(\Lambda_b \to \Lambda_c \tau \bar\nu)/\d q^2$ spectrum as a blue band.  

\begin{table}[tb]
\renewcommand{\arraystretch}{1.2}
\begin{tabular}{c|rrrrrr}
\hline\hline \\[-10pt]
& $\zeta'$ & $\zeta''$ &  $\hat b_1$	 & $\hat b_2$ & $m_b^{1S}$ & $\delta m_{bc}$  \\	\hline
$\zeta'$	&$ 1.00$&$ -0.94$&$ -0.14$&$ 0.11$&$ 0.11$&$ -0.01$\\ 
$\zeta''$	&$ -0.94$&$ 1.00$&$ 0.13$&$ -0.02$&$ -0.10$&$ 0.00$\\ 
$\hat b_1$  &$ -0.14$&$ 0.13$&$ 1.00$&$ 0.10$&$ -0.21$&$ 0.10$\\ 
$\hat b_2$  &$ 0.11$&$ -0.02$&$ 0.10$&$ 1.00$&$ -0.63$&$ 0.05$\\ 
$m_b^{1S}$  &$ 0.11$&$ -0.10$&$ -0.21$&$ -0.63$&$ 1.00$&$ -0.00$\\ 
$\delta m_{bc}$ &$ -0.01$&$ 0.00$&$ 0.10$&$ 0.05$ &$ -0.00$&$ 1.00$\\ 
\hline\hline
\end{tabular}
\caption{Correlation matrix of the HQET parameters determined from the fit to
the LHCb measurement and the LQCD form factors.}
\label{tab:fitcov}
\end{table}

\section{Conclusions}
\label{sec:concl}

Measurement of $\Lambda_b \to \Lambda_c \ell\bar\nu$ decays will play an important role in elucidating the tantalizing hints of new physics in the measurements of $R(D^{(*)})$, and refining our understanding of determinations of the CKM element $|V_{cb}|$.
We derived new model independent predictions for these decays, and found that fitting the LHCb data for $\d\Gamma(\Lambda_b\to\Lambda_c\mu\bar\nu)/\d q^2$ substantially reduces the uncertainty of the SM prediction for $R(\Lambda_c)$. We obtained
\begin{align}
 \RL = 0.324 \pm 0.004 \,,
\end{align}
by combining the lattice information with the measured spectrum. This produces the most precise prediction of $\RL$ to date, 
significantly improving the precision over the lattice QCD prediction, $\RL = 0.3328 \pm 0.0070 \pm 0.0074$~\cite{Detmold:2015aaa}.
This large improvement arises because the experimental data constrain combinations of form factors relevant for the prediction of $\RL$.

Using the lattice QCD form factor calculations, we performed new tests of heavy quark symmetry, determining $\lqcd^2/m_c^2$ corrections to an exclusive decay, without any model dependent assumptions, for the first time. The HQET expansion at order $\lqcd^2/m_c^2$ appears well behaved, and we find good agreement between lattice QCD and HQET predictions.
More details and extensions of these results including new physics
contributions will be presented elsewhere~\cite{baryon:long}.

\begin{acknowledgments}

We thank the Aspen Center of Physics, supported by the NSF grant 
PHY-1607611, where parts of this work were completed. 
We thank HG for inspiring color choices to scoop FT. FB thanks JMB for seven wonderful years together.
FB and WS were supported by the DFG Emmy-Noether Grant No.\ BE~6075/1-1.   ZL was supported in part by the U.S.\ Department of Energy under
contract DE-AC02-05CH11231. The work of DR was supported in part by NSF grant PHY-1720252.
\end{acknowledgments}

\bibliographystyle{apsrev4-1}
\bibliography{baryon}

\end{document}